# On dependence of the source deactivation factor on the magnitude of main earthquake shock

A.V. Guglielmi, A.D. Zavyalov, O.D. Zotov
*Schmidt Institute of Physics of the Earth RAS,Moscow, Russia*
*guglielmi@mail.ru, zavyalov@ifz.ru, ozotov@inbox.ru*

## Abstract

We have discovered an inverse relationship between the earthquake source deactivation coefficient and the magnitude of the main shock. Recently, this pattern was confirmed through an analysis of earthquakes in the Central Himalayas (Poudel et al., 2026). On the one hand, independent confirmation of a previously unknown pattern strengthens our confidence in the relevance of the phenomenological theory of aftershocks, upon which we have based our method for processing and analyzing aftershocks. On the other hand, Poudel et al. (2026) presented our method in a distorted manner. Poudel et al. mistakenly believe that our method is based on a modified Omori law. Both circumstances provide an occasion and a basis for a simple and clear explanation of the essence of our approach to the study of aftershocks. First and foremost, our approach to the problem is diametrically opposed to Omori's approach. Omori's law and its modification, proposed by Hirano and Utsu, describe the decay of aftershocks. In contrast, we formulated and solved the inverse problem. The essence of the inverse problem lies in the need to calculate the source deactivation coefficient based on aftershock observation data, without making any prior assumptions regarding the law governing the decay of aftershocks. A crucial concept in our theory is the rather complex notion of the proper time of a source. Proper time differs radically from astronomical time. The paper describes the continuous and discrete representations of proper time. This paper is methodological in nature and may prove useful to researchers studying the dynamics of an earthquake source following the formation of a main rupture within the rock mass.

## 1. Introduction

It has been established, both theoretically and experimentally, that the source deactivation coefficient $\sigma$ during the Omori epoch decreases as the magnitude $M$ of the earthquake's mainshock increases [Guglielmi, Zotov, 2021]. Figure 1 shows the relationship $\sigma(M)$. The coefficient of determination is 0.82.

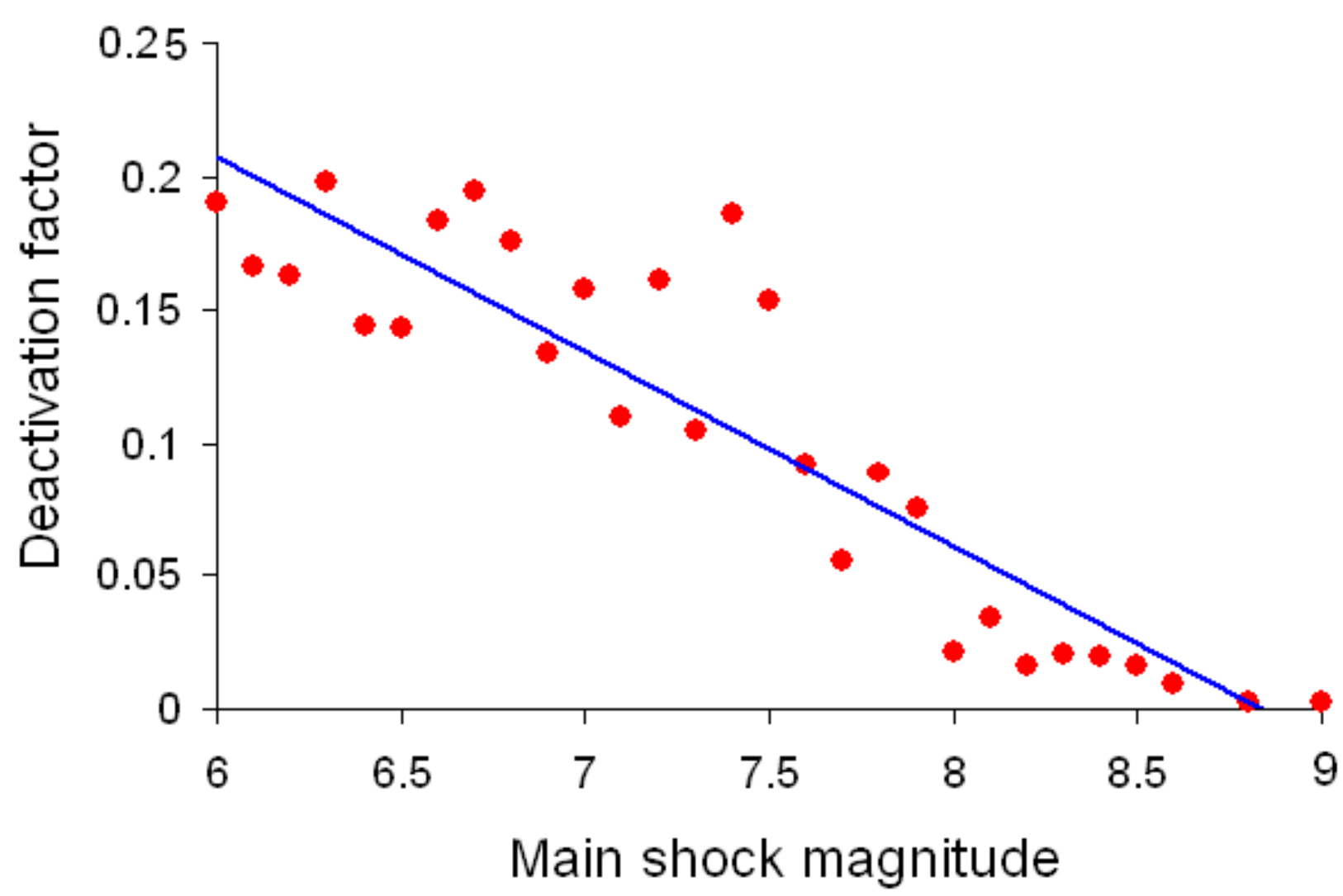

Fig. 1. Dependence of the source deactivation coefficient $\sigma$ in the Omori epoch on the magnitude of the main shock.

The source deactivation coefficient was calculated using the formula

$$\sigma(t) = \frac{d}{dt}\langle T(t) \rangle, \qquad (1)$$

where $T(t) = 1/n(t)$, $n(t)$ is the aftershock frequency, and the angle brackets denote optimal smoothing of the function $T(t)$. Formula (1) is derived within the framework of a phenomenological theory of aftershocks, which we developed in our papers [Guglielmi, 2016, 2017; Guglielmi, Zavyalov, 2018; Zotov et al., 2018; Zavyalov et al., 2020; Guglielmi, Zotov, 2021, 2024; Zavyalov, Zotov, 2021; Zavyalov et al., 2022; Guglielmi et al., 2023, 2024, 2025; Zotov, Guglielmi, 2025a, 2025b; Guglielmi, Zotov, 2026].

This article focuses on a methodology we have proposed for studying the relaxation of an earthquake source following the main shock. The impetus for writing this article was a recently [Poudel et al., 2026]. In it, the authors thoroughly investigated aftershocks in the Himalayas using our methodology and convincingly confirmed that the deactivation coefficient decreases as the magnitude of the mainshock increases. This is remarkable, as it reinforces confidence in the relevance of the phenomenological theory of earthquakes. At the same time, the authors of the aforementioned work believe that our method for studying aftershocks is based on a modified Omori law. Meanwhile, our approach to the problem is diametrically opposed to the traditional approach, which was successfully developed by Omori, Harano, Utsu, and many other researchers. We wish to demonstrate that our methodology does not in any way replace traditional methods for studying aftershocks, rather, it complements them and significantly expands the possibilities for the experimental discovery of new properties and patterns of earthquakes.

## 2. Modified Omor's Law (MOL)

First and foremost, it must be said that we hold Fusakichi Omori, a student of John Milne, in the highest regard. Both prominent scientists laid the experimental and theoretical foundations of modern seismology. We have dedicated papers to the memory of Omori [Guglielmi, Zavyalov, 2018; Guglielmi et al., 2024] and have consistently highlighted the significance of Omori's law in the physics of aftershocks.

The classical Omori law was established at the end of the century before last [Omori, 1894]. But science does not stand still. Living science evolves, and as early as the beginning of the last century, Hirano proposed modifying Omori's law as follows:

$$n(t) = k / t^{p}. \qquad (2)$$

Here,$n(t)$ is the aftershock frequency, $t$ is the time elapsed since the main shock, and $k$, $p$ are the fitting parameters [Hirano, 1924]. All the quantities specified here are positive. For $p = 1$, formula (2) describes the classical Omori law. For an arbitrary positive value of $p$, formula (2) is called the modified Oiori law (abbreviated as MOL). MOL is widely used in experimental and theoretical studies of aftershocks (see e.g. [Utsu, 1961, 1962, 1971; Ogata, 1988; Molchan, Dmitrieva, 1992; Wang,1994; Utsu et al., 1995; Kisslinger, 1996; Ogata, Guo, 1997; Lombardi, 2002; Ogata, Zhuang, 2006; Holschneider et al., 2012; Shcherbakov et al., 2018; Rodrigo, 2021; Salinas-Martínez et al., 2023]). The significance of formula (2) lies primarily in the fact that it provides a compact quantitative description of the aftershock sequence.

However, in the context of this paper, it is particularly significant that MOL is fundamentally incompatible with a methodology based on describing an earthquake source using the deactivation coefficient. Thus, the view held by the authors of [Poudel et al., 2026] that our methodology for studying the $\sigma(M)$ relationship is based on MOL is incorrect.

The situation is as follows. From formula (2), taking (1) into account, it follows that

$$\sigma(t) = (p/k)t^{p-1}. \quad (3)$$

We see that MOL predicts a monotonic dependence of the deactivation coefficient on time. For $p<1$ ($p>1$), the function $\sigma(t)$ monotonically decreases (increases) over time. This property of MOL is inconsistent with the extensive experience gained from studying earthquake source relaxation using the methodology we proposed. A non-monotonic dependence $\sigma(t)$ was observed in all the aftershocks we studied, without exception.

The classical Omori law ($p=1$) predicts that the deactivation coefficient of an earthquake source remains constant. In practice, however, the equality $\sigma = \text{const}$ holds true only for a limited span of time, which we have termed the Omori epoch. For example, in the case of the Tohoku earthquake, the Omori epoch lasted for a month following the main shock. Following the end of the Omori epoch, the value $\sigma(t)$ fluctuated chaotically over the course of several months. In other words, Omori's law is not holistic. We will describe the holistic law of aftershock decay in the next section of the paper.

## 3. Law of Aftershock Decay (LAD)

The methodology we propose is novel and significant, but it may appear unnecessarily complex. Therefore, we will outline our approach concisely and, as far as possible, in the simplest terms.

As a basis for a phenomenological theory of aftershocks, we shall adopt the concept of earthquake source as a dynamic system whose state is characterized by deactivation coefficient. Our problem is the study of the dynamics of an earthquake source undergoing relaxation following the main shock. We view aftershocks as a source of information about the relaxation of the focal zone. The equation for aftershock evolution follows logically from a simple postulate: The aftershock rate $n(t)$ is a smooth, slowly decaying function of time. The condition of slowness implies that the relative change in the aftershock frequency over a time interval $T = 1/n$ is much less than unity. Mathematically, this is expressed by the strict inequality

$$n \gg \left| \frac{d}{dt} \ln n \right| . \tag{4}$$

Let $\sigma$ be a small dimensionless parameter, $0 < \sigma \ll 1$. We can rewrite (4) as

$$\frac{dn}{d\tau} + n^2 = 0 , \; \tau(t) = \int_0^t \sigma(t')dt' . \tag{5}$$

We shall call the quantity $\tau$ the proper time of the earthquake source. We see that, when using proper time as the chronometric scale, the evolution equation for aftershocks takes on an exceptionally simple form. The proper time of the source flows more slowly than astronomical time. The second important property of proper time is that it may flow non-uniformly, due to the non-stationary of the conditions at earthquake source.

Let us perform a substitution for the dependent variable $n \to T = 1/n$. In that case, the solution to the Cauchy problem for the evolution equation takes on an exceptionally simple form when events are timed according to the proper time of the earthquake source:

$$T(\tau) = T_0 + \tau . \tag{6}$$

Here, $T_0 = T(0)$ is the initial condition. We have obtained a holistic law of aftershock decay (referred to as LAD). Let us state the LAD formulation in words: the astronomical waiting time for the next aftershock increases linearly with the passage of proper time.

We cannot use solution (6) to analyze aftershocks until we specify the deactivation coefficient. The authors of the paper [Poudel et al., 2026] mistakenly believe that MOL provides us with such an opportunity. However, we pointed out in the previous section that MOL contradicts experiment. In fact, at present, we are not aware of any mechanisms that lead to non-monotonic variations in the deactivation coefficient of earthquake source. Under these conditions, we will shift our research strategy to the exact opposite. We will forgo solving the Cauchy problem and instead seek a solution to the inverse problem. The inverse problem consists of calculating the deactivation coefficient based on data regarding the observed activity of aftershocks. A correct solution to the inverse problem is given by formula (1).

Experience in the practical solution of the inverse problem indicates that the deactivation coefficient is variable. This circumstance was not taken into account in the article [Poudel et al., 2026]. It may be assumed, however, that the results of the aforementioned work date from the Omori epoch.

## 4. Discussion and Conclusions

Generally speaking, describing aftershock activity using a continuous function $n(t)$ is an idealization. After all, aftershocks are a discrete sequence of seismic events. The phenomenological theory of aftershocks allows for a transition from a continuum-based to a discrete representation of LAD. Let us discuss the result of such a transition.

Let us number the aftershocks using the sequence of natural numbers $i = 1, 2, 3, ..., N$. Let us determine the excitation times of each aftershock $t_i$ from the earthquake catalog.

Let us introduce a coordinate plane ($x$, $t$) and plot the points $(i,t_i)$ on it. Let us introduce the notations $T_i = t_i - t_{i-1}$, $\vartriangle T_i = T_i - T_{i-1}$, and $\tau_i = \tau(t_i)$. The transition from a continuous representation to a discrete one is not unique. If formula (6) is used, then

$$\Delta T_i = \int_{t_i-1}^{t} \sigma(t)dt . \qquad (7)$$

Using the mean value theorem to evaluate the integral, we obtain the recurrence relation

$$T_i = T_{i-1} / (1 - \bar{\sigma}_i),\ i \geq 3 . \qquad (8)$$

The quantities $\bar{\sigma}_i$ are unknown to us, and, furthermore, condition $0 < \bar{\sigma}_i \ll 1$ which is necessary for the applicability of formula (6) may be violated. This consideration suggests the need to impose a condition requiring the deactivation coefficient to change slowly over a time interval on the order of $1/n$:

$$n \gg \left| \frac{d}{dt} \ln \sigma \right| . \qquad (9)$$

Condition (9), like condition (4), is necessary for the concept of the instantaneous frequency of aftershocks introduced into seismology through the work of Omori, Hirano, and Utsu to have a meaningful interpretation.

Condition (9) is satisfied in the Omori epoch. For the Omori epoch, we replace $\bar{\sigma}_i$ with $\sigma$ in (8) and find that, on average, the quantities $T_i$ form a geometric progression:

$$T_i = T_{i-1}(1+\sigma),\ i \geq 3 . \qquad (10)$$

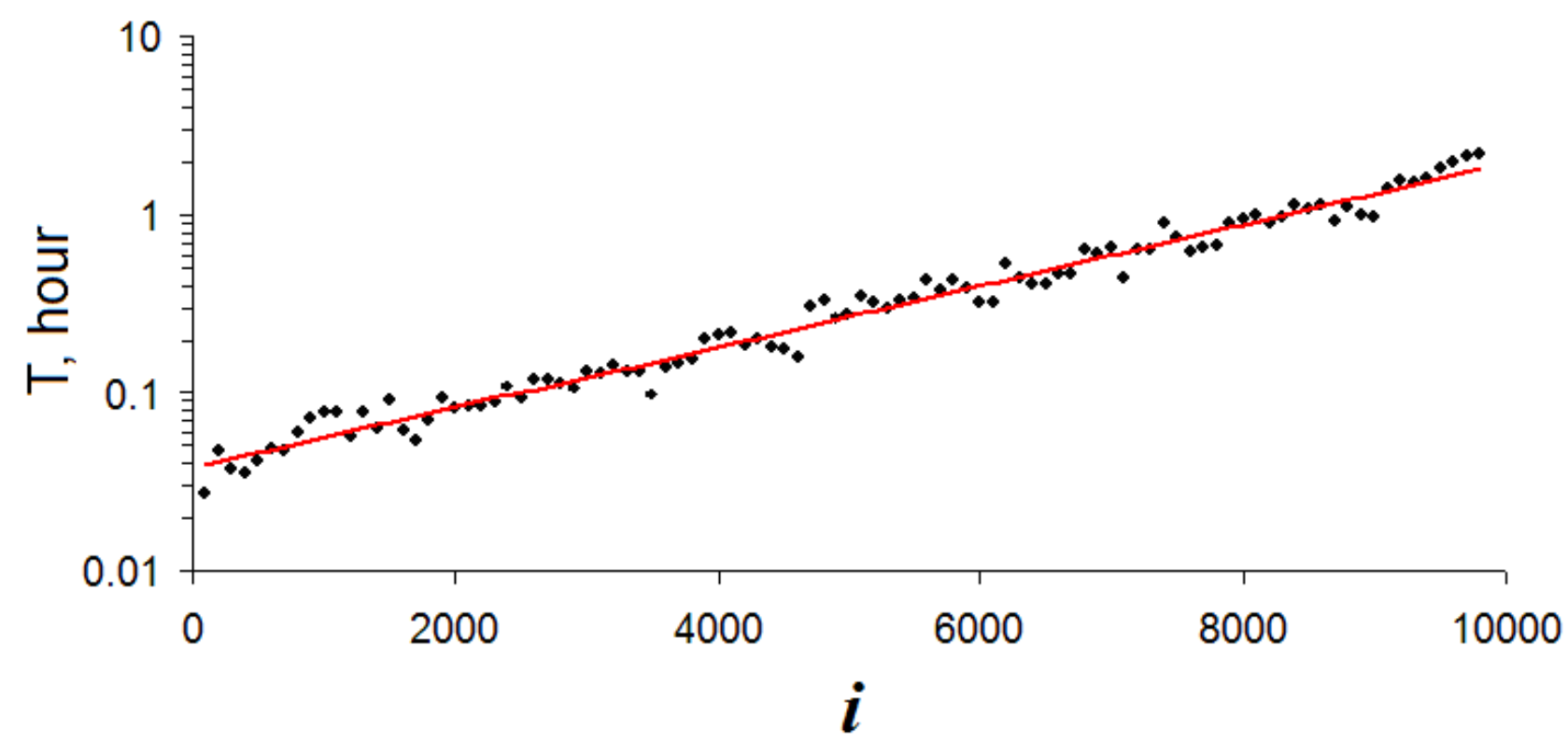

Fig. 2. Average waiting time for the next aftershock. The straight line passes through points of a geometric progression that approximates the experimental data points.

Figure 2 clearly illustrates our assertion [Zotov, Guglielmi, 2025]. The SCEC catalog of earthquakes in Southern California was used to create the figure. (https://scedc.caltech.edu). The main shock of the earthquake, with a magnitude of $M = 6.7$, occurred on January 17, 1994. Each point indicates the average waiting time, calculated from 100 consecutive aftershocks.

To conclude the discussion, we shall describe an alternative representation of proper time distinct from the $\tau$ -representation outlined in the previous section of the paper, and refer to it as the $x$-representation. Let us define the synchronization function $t(x)$ as a smooth function that approximates the experimental points $(i, t_i)$ in the $(x, t)$ coordinate plane. Due to the irreversibility of time, there also exists an inverse function $x(t)$. Let us present an example of a synchronization function for the first N = 4500 aftershocks of the Tohoku earthquake [Guglielmi, Zotov, 2026]

$$t(x) = 7.2 \exp\left[0.0014(x-1)\right] \text{, h.} \tag{11}$$

The coefficient of determination is close to unity: $R^2 = 0.97$.

Let us note that the graph of a geometric progression is an exponential function. It is easy to verify that, in this case, the *x*-representation leads us to a linear evolution equation for aftershocks

$$\frac{dn}{dx} + \sigma n = 0. \tag{12}$$

It follows from this that aftershocks decay exponentially in the proper-time representation

$$n(x) = n_1 \exp\left[-\int_1^x \sigma(x')dx'\right]. \tag{13}$$

The solution to the inverse problem is given by the formula

$$\sigma(x) = \frac{d}{dx}\langle \ln t(x) \rangle. \tag{14}$$

Calculation using formula (14) for the Omori epoch of the Tohoku earthquake yields a value of $\sigma = 0.0014$, with a coefficient of determination $R^2 = 0.97$.

We have the opportunity to perform two critical checks of the deactivation coefficient value that has been found. First, $\sigma$ for the Tohoku earthquake can be calculated using formula (1), which we derived in the proper-time $\tau$ –representation. The result proved favorable: $\sigma = 0.0014$, with a coefficient of determination $R^2 = 0.82$.

A second possibility is provided by an analysis of the work of [Poudel et al., 2026]. Based on observations in the Central Himalayas, an empirical relationship

$$\sigma(M) = 0.11 - 0.01M \tag{15}$$

has been found for $M \leq 7.6$. We will have to perform a significant extrapolation, since the value $\sigma = 0.0014$ was obtained by us at $M = 9.1$. Nevertheless, the result turns out to be quite interesting. According to formula (15), the deactivation coefficient for the Tohoku source is $\sigma = 0.0019$, which is virtually identical to the result obtained from independent calculations using both formula (1) and formula (14).

Let us present a summary of our detailed analysis regarding the dependence of earthquake source deactivation factor on the magnitude of the main shock:

1. Contrary to the view expressed by the authors of the paper [Poudel et al., 2026], the methodology we propose for investigating source dynamics is not based on the modified Omori's law, but rather on solving an inverse problem for the evolution equation that describes the law of aftershock decay.

2. The most important concepts in the phenomenological theory of aftershocks are the deactivation coefficient and the proper time of a source undergoing relaxation following the main shock.

3. The reckoning of proper time is possible under both continuous and discrete representations of an aftershock sequence, with the two representations being consistent with each other.

What mechanism is at work within an earthquake source, leading to a geometric progression of seismic shocks during the Omori epoch? How does a simple arithmetic theory determine the chronology of aftershocks? We cannot answer these questions. We simply pay special attention to them. It is doubtful that the problem can be solved within the framework of the phenomenology we have presented here, or in light of mathematical catastrophe theory [Guglielmi, 2015]. Understanding will most likely be achievable based on a foundation of physical concepts regarding the earthquake source. We place particular hope in the Umov energy flows within the lithosphere, which, by all indications, manifest as the convergence of foreshocks and the divergence of aftershocks associated with a major earthquake [Guglielmi, Zotov, 2024].

*Acknowledgments*. We thank F.Z. Feygin and A.S. Potapov for their interest in our work and their steadfast support. This work was supported by the Ministry of Science and Higher ducation of the Russian Federation within the framework of the state assignment of the O.Yu. Schmidt Institute of Physics of the Earth of the Russian Academy of Sciences.